## *P*-type conductivity in cubic GaMnN layers grown by molecular beam epitaxy.

S V Novikov[1,2], K W Edmonds[1], A D Giddings[1], K Y Wang[1], C R Staddon[1], R P Campion[1], B L Gallagher[1] and C T Foxon[1,3]

[1]*School of Physics and Astronomy, University of Nottingham, Nottingham, NG7 2RD, UK*

[2)] On leave from the Ioffe Physical-Technical Institute, St.Petersburg, Russia
[3)] Corresponding author – Fax: +44-115-9515180, Phone: +44-115-9515138,
e-mail: C.Thomas.Foxon@nottingham.ac.uk

## Abstract

Cubic (zinc-blende) $Ga_{1-x}Mn_xN$ layers were grown by plasma-assisted molecular beam epitaxy on GaAs (001) substrates. Some of the structures also contained cubic AlN buffer layers. Auger Electron Spectroscopy and Secondary Ion Mass Spectroscopy studies clearly confirmed the incorporation of Mn into cubic GaN, the Mn being uniformly distributed through the layer. X-ray diffraction studies demonstrated that the Mn-doped GaN films are cubic and do not show phase separation up to a Mn concentrations of $x<0.1$. *P*-type conductivity for the cubic $Ga_{1-x}Mn_xN$ layers was observed for a wide range of the Mn doping levels. The measured hole concentration at room temperature depends non-linearly on the Mn incorporation and varies from $3x10^{16}$ to $5x10^{18}$ $cm^{-3}$.

III-V ferromagnetic semiconductors (FS) offer a unique combination of electronic, magnetic, and optical properties, which promise a revolution in data storage and processing technologies, and hold prospects for quantum computation [1,2]. In these systems, ferromagnetism is mediated by exchange interactions between localised *d*-electrons and itinerant charge carriers (holes). The ferromagnetic transition temperature $T_C$ depends on the densities of both the magnetic impurities *x* and holes *p* [3]. It is therefore essential for spintronic applications to have films with high *p*-type conductivity and a high density of magnetic impurities. Ideally this is achieved using an appropriate magnetic impurity, which also provides the free holes as is the case for Mn in GaMnAs. For example, for the GaMnAs system, the best $T_C$ reported by our group for single layers without co-doping is 159K [4], however with co-doping significantly higher values for $T_C$ have been reported (for example 280K [5]). However, technological applications require a $T_C$ above room temperature, which may require a different material system.

There are theoretical predictions for room temperature carrier-mediated ferromagnetism in *p*-type $Ga_{1-x}Mn_xN$ (x>0.05 and $p>3.5\times10^{20}$ cm$^{-3}$), both from effective Hamiltonian [3] and density-functional [6] calculations. Such predictions have stimulated experimental studies of this system and different values for $T_C$ from 350K to 940K have been reported for hexagonal (Wurtzite) GaMnN [7-9]. However, there is no clear correlation between the reported carrier density and magnetic properties for GaMnN. Experimentally, most authors report that their GaMnN films are *n*-type, apparently ruling out a carrier-mediated origin for the ferromagnetism since conduction electrons will interact only weakly with the magnetic impurities. However, there have been a few reports of p-type conductivity in wurtzite GaMnN layers (see for example [10]). It has been theoretically and experimentally suggested that the Mn level is too deep in the GaN band gap to be an effective acceptor [7-9], explaining why *p*-type conduction is not generally observed. Therefore, the reported ferromagnetism has been ascribed to undetected metallic secondary phases [11,12].

Most of the GaMnN experimental work to date has been carried out on wurtzite materials, but there are very few reports for similar studies on cubic (zinc-blende) GaMnN [13]. Theoretical studies suggest that the Mn level in zinc-blende GaN will also be deep in the gap [14], similar to the situation for wurtzite material. However, experimentally there are reports that the carbon level in zinc-blende GaN is much shallower [15] than the corresponding level for wurzite GaN [16]. Thus carbon is an effective acceptor in zinc-blende GaN [15], but produces insulating films in wurtzite GaN [16].

Based on the above considerations, the aim of the current work was to check experimentally whether *p*-type conductivity can be achieved in cubic $Ga_{1-x}Mn_xN$ and to study the incorporation of Mn in cubic GaN for a wide range of the Mn concentrations.

Undoped cubic GaN films and cubic GaMnN layers were grown on semi-insulating GaAs (001) substrates by plasma-assisted molecular beam epitaxy (PA-MBE) using arsenic as a surfactant to initiate the growth of cubic phase material [17]. The substrate temperature was measured using an optical pyrometer. Growth temperatures from 450 to 680$^o$C were used. The active nitrogen for the growth of the group III-nitrides was provided by an CARS25 RF activated plasma source. Prior to the growth of active nitride layers, a GaAs buffer layer was grown on the GaAs substrate in order to improve the properties of the $Ga_{1-x}Mn_xN$ layers. In some samples an undoped cubic GaN (~150nm thick) buffer layer followed by a cubic AlN buffer layer (50-150 nm thick) was introduced between GaAs layer and cubic $Ga_{1-x}Mn_xN$ layer in order to ensure electrical isolation. The Mn concentration in the films was set using the in-situ beam monitoring ion gauge, assuming a sticking coefficient for Mn of unity.

Samples were studied in-situ using the reflection high-energy electron diffraction (RHEED) and after growth ex-situ measurements were performed using Auger Electron Spectroscopy (AES), Secondary Ion Mass Spectroscopy (SIMS), X-ray diffraction (XRD) and Hall effect measurements.

RHEED studies during the growth of the GaAs buffer layer showed the conventional $2 \times 4$ reconstruction. RHEED patterns observed during and after the growth of $Ga_{1-x}Mn_xN$ layers obtained under N-rich conditions showed the usual spotty pattern seen for nitride semiconductors. For similar layers grown under Ga-rich conditions a streaky RHEED pattern with a $2 \times 2$ reconstruction is observed during growth and after growth. In-situ RHEED studies showed that all the $Ga_{1-x}Mn_xN$ and AlN buffer layers have cubic symmetry.

X-ray studies of the films confirm that the GaMnN and AlN films are cubic and epitaxial with respect to the GaAs buffer layer, as shown in figure 1. The peak at $\sim 40^o$ corresponds to cubic GaN, the remaining two peaks come from the GaAs substrate and GaAs epitaxial layer. No evidence for hexagonal GaMnN, GaN or AlN inclusions was observed within the sensitivity limit of the X-ray system. For Mn concentration $x>0.1$, evidence for phase separation is observed with the possible formation of $GaMn_3N$ or $Mn_4N$ inclusions.

Auger measurements clearly demonstrated the existence of the Mn atoms in the $Ga_{1-x}Mn_xN$ layers. The ratio of the Mn/Ga AES signals increased with increasing Mn flux during MBE. The $Ga_{1-x}Mn_xN$ layer composition estimated from this Mn/Ga ratio measured by AES was close to the composition estimated from the in-situ beam monitoring ion gauge in the MBE system.

SIMS studies also confirmed the incorporation of Mn in the cubic GaMnN films as shown in figure 2. The Mn SIMS concentration is determined using a Mn implantation standard in GaN. The Mn concentration in the GaMnN film (figure 2) is uniform and is approximately equal to $4.5\times10^{20}$ $cm^{-3}$, which gives a value of $x\sim0.01$. This is comparable to the figure estimated from the in-situ beam monitoring ion gauge. The Mn concentration decreases at the GaMnN/GaAs interface at the same rate as the N concentration to the background level for the SIMS system. This suggests that both N and Mn slopes are due to the non-uniform sputtering process in the SIMS profiling. This indicates that there is no significant diffusion of Mn into the underlying GaAs buffer layer. SIMS studies also show that the Mn concentration increases with the Mn flux used during growth, from $x=0$ to $x>0.1$. The SIMS data also shows a significant concentration of As in the GaMnN layer. Using two different secondary ions, the estimated As concentration is slightly different in the GaMnN films due to different As locations in the GaN lattice as discussed in detail elsewhere [18], but is identical in the GaAs buffer layer and substrate as expected.

Hall effect measurements show that *p*-type conductivity is observed for a wide range of Mn concentrations in cubic GaMnN, whereas undoped GaN films are *n*-type. The measured hole concentration at room temperature depends non-linearly with the Mn incorporation and varies from $3x10^{16}$ to $5x10^{18}$ $cm^{-3}$. This *p*-type conductivity can be observed for films grown at different temperatures in the range 450-680$^0$C. Figure 3 shows the Hall resistance measurements on a cubic $Ga_{1-x}Mn_xN$ sample with a nominal value $x=0.07$, together with the carrier densities obtained from these measurements as a function of temperature. The sample clearly shows *p*-type conduction. The temperature dependence of the hole concentration deduced from the Hall data is shown in the insert. The measured carrier concentration $p$ increases with increasing temperature consistent with thermal ionisation, with $p$ about $10^{18}$ $cm^{-3}$ at room temperature. The apparent ionisation energy, $E_A$, estimated from figure 3, assuming uncompensated material, is ~54meV for Mn in cubic $Ga_{1-x}Mn_xN$.

It is well known that Mn-doped GaAs is *p*-type. SIMS data presented above indicates that there is no significant diffusion of Mn into the GaAs buffer layer. However, to completely eliminate the possibility that the *p*-type conductivity in the GaMnN layers is due to Mn diffusion into the underlying GaAs layer, we have grown additional structures with a thick (~150nm) undoped cubic GaN buffer layer followed by a thick (50-150nm) insulating cubic AlN buffer layer between the GaAs and cubic GaMnN layers. The electrical properties of the films grown without and with cubic GaN/AlN buffer layers of different thickness have similar hole densities and all show *p*-type conductivity. Therefore the *p*-type conductivity must be associated with Mn in the cubic GaMnN layers.

Having established that the electrical conductivity is associated with the GaMnN layers, there may be alternative explanations for the observed p-type behaviour. The three potential dopants responsible for p-type conductivity in cubic GaMnN include Mg, C and Mn. There is no possible source of Mg in our system. The second possibility is that unintentional C-doping of the cubic GaMnN films is responsible for the p-type conductivity. However, SIMS studies show that the unintentional O and C levels in the GaMnN films are similar to those observed in undoped n-type GaN films, O always being higher than C. Furthermore the unintentional doping levels of O and C do not change systematically with Mn concentration, whereas the electrical properties do change significantly, this will be discussed in more detail elsewhere. Therefore the *p*-type conductivity must be due to the Mn in the cubic GaMnN layers.

The electrical, optical and magnetic properties of the cubic GaMnN films are now being investigated in detail and will be reported elsewhere. The measured hole concentration does not change monotonically with Mn content, which is now the subject of our more detailed studies.

In summary, cubic (zinc-blende) $Ga_{1-x}Mn_xN$ films showing *p*-type conduction have been grown by PA-MBE on GaAs substrates for a wide range of Mn concentrations, growth temperature and V:III ratio both with and without a cubic AlN buffer layer. Our Hall effect measurements experimentally demonstrate that it is possible to obtain *p*-type conductivity in cubic GaMnN films. This opens the possibility to obtain carrier-mediated ferromagnetism at room temperature in this materials system.

## Acknowledgements

This work was undertaken with support from EPSRC (GR/R46465) and by the EC (FENIKS G5RD-CT-2001-00535). The authors would also like to acknowledge the contributions of B.Ja. Ber and A.P. Kovarsky for the SIMS studies and J. Cibert and H. Mariette for helpful discussions on the growth of GaMnN.

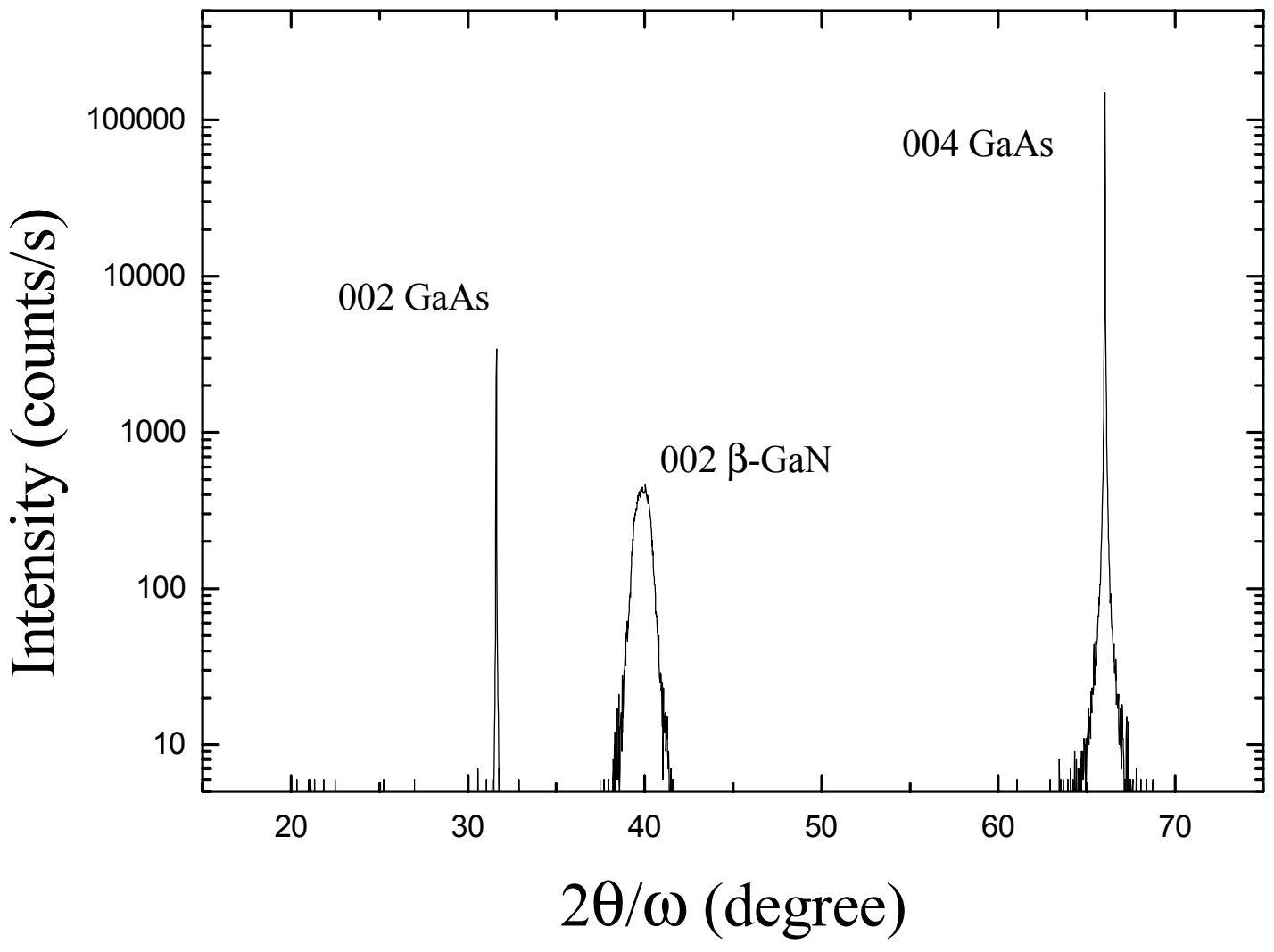

Figure 1. X-ray data for a $Ga_{1-x}Mn_xN$/GaAs heterostructure with x~0.05; the GaAs layer is about 150 nm thick and the $Ga_{1-x}Mn_xN$ is about 300 nm thick.

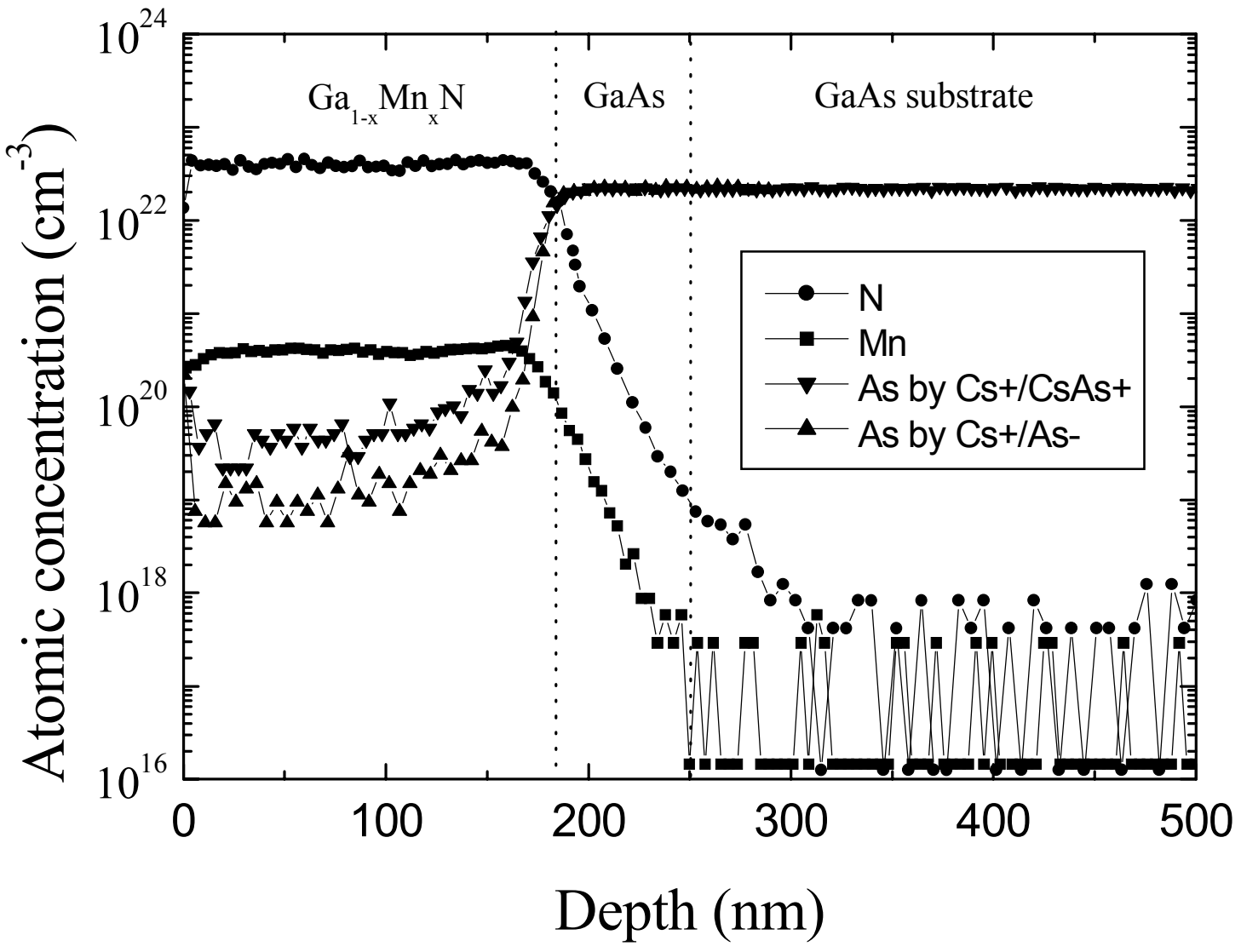

Figure 2. SIMS profiles for Mn, N and As in a $Ga_{1-x}Mn_xN$/GaAs heterostructure with x~0.01. The positions of the interface between the GaMnN/GaAs-epilayer/GaAs-substrate are indicated on the figure.

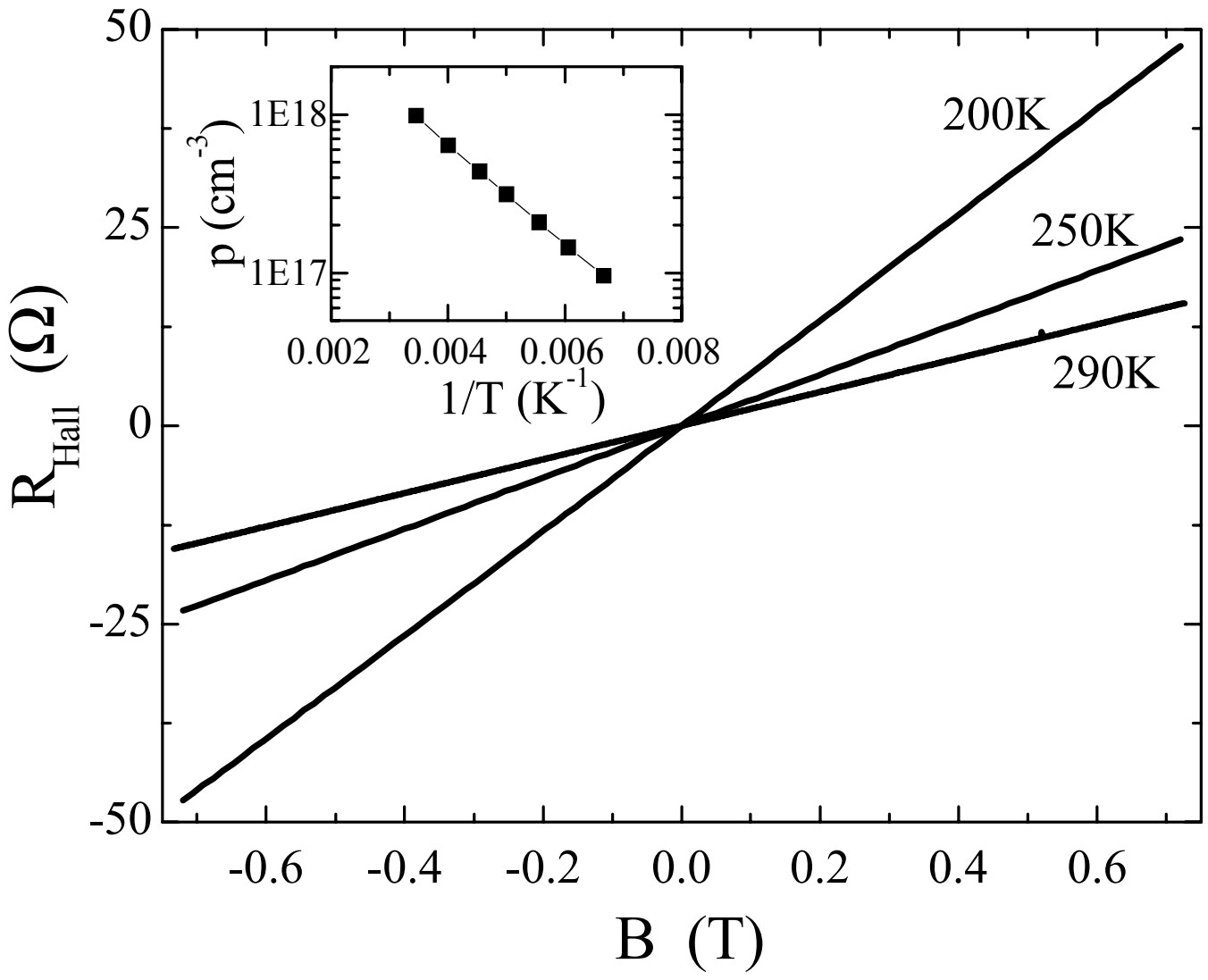

Figure 3. Hall effect data for a $Ga_{1-x}Mn_xN$ film with x~0.07. The temperature dependence of the hole concentration deduced from the Hall data is shown in the insert.